\documentclass[twocolumn,english,aps,prl,showpacs,amsmath,amssymb,superscriptaddress]{revtex4}
\usepackage{graphicx}
\usepackage{amssymb}
\usepackage{natbib}
\usepackage{color}
\usepackage{amsmath}

\begin{document}

\title{Spin excitation anisotropy in the paramagnetic tetragonal phase of BaFe$_2$As$_2$}

\author{Yu Li}
\affiliation{Department of Physics and Astronomy,
Rice University, Houston, Texas 77005, USA}

\author{Weiyi Wang}
\affiliation{Department of Physics and Astronomy,
Rice University, Houston, Texas 77005, USA}

\author{Yu Song}
\affiliation{Department of Physics and Astronomy,
Rice University, Houston, Texas 77005, USA}

\author{Haoran Man}

\affiliation{Department of Physics and Astronomy,
Rice University, Houston, Texas 77005, USA}

\author{Xingye Lu}
\affiliation{Center for Advanced Quantum Studies and Department of Physics, Beijing Normal University, Beijing 100875, China}

\author{Fr$\rm \acute{e}$d$\rm \acute{e}$ric Bourdarot}
\affiliation{Institut Laue Langevin, 71 Avenue des Martyrs, 38042 Grenoble, France}

\author{Pengcheng Dai}
\email{pdai@rice.edu}
\affiliation{Department of Physics and Astronomy,
Rice University, Houston, Texas 77005, USA}
\affiliation{Center for Advanced Quantum Studies and Department of Physics, Beijing Normal University, Beijing 100875, China}

\date{\today}
\pacs{74.70.Xa, 75.30.Gw, 78.70.Nx}

\begin{abstract}
We use neutron polarization analysis to study temperature dependence of the spin excitation anisotropy in BaFe$_2$As$_2$, which has a tetragonal-to-orthorhombic structural distortion at $T_s$ 
and antiferromagnetic (AF) phase transition at $T_N$ with ordered moments along the orthorhombic $a$-axis 
below $T_s\approx T_N\approx 136$ K. In the paramagnetic tetragonal state at 160 K,
spin excitations are isotropic in spin space with $M_a=M_b=M_c$, where $M_a$, $M_b$, and $M_c$ are spin excitations polarized along
the $a$, $b$, and $c$-axis directions of the orthorhombic lattice, respectively.  
On cooling towards $T_N$, significant spin excitation anisotropy with $M_a>M_b\approx M_c$ develops below 3 meV
 with a diverging $M_a$ at $T_N$.  The in-plane spin excitation anisotropy
in the tetragonal phase of BaFe$_2$As$_2$ is similar to those seen in the tetragonal phase of its electron and hole-doped superconductors, suggesting that spin excitation anisotropy is a direct probe of doping dependence of spin-orbit coupling and its connection to superconductivity in iron pnictides.
\end{abstract}

\maketitle

The iron pnictide superconductors have a rich phase diagram including an orthorhombic lattice distortion associated with ferro-orbital order and nematic phase, 
antiferromagnetic (AF) order, and 
superconductivity \cite{hosono,stewart,dai,fisher,RMFernandes2014,AEBohmerCRP}. 
In the undoped state, a parent compound of iron pnictide superconductors BaFe$_{2}$As$_{2}$ 
forms stripe AF order at $T_{N}$ near a tetragonal-to-orthorhombic structural transition
temperature $T_s$ [Fig. 1(a)] \cite{qhuang,kim2011,Wilson10}. Superconductivity can be induced by partially replacing Ba by K in
BaFe$_{2}$As$_{2}$ to form hole-doped Ba$_{1-x}$K$_{x}$Fe$_{2}$As$_{2}$ or by partially replacing
Fe by $TM$ ($TM=$Co, Ni) to form electron-doped BaFe$_{2-x}TM_{x}$As$_{2}$ \cite{hosono,stewart,dai}.
Although much attention has been focused on understanding the 
interplay between magnetism and superconductivity in these materials \cite{hosono,stewart,dai}, a more subtle and much less explored facet involves 
the effect of spin-orbit coupling (SOC) \cite{Borisenko}, which translates anisotropies in
real space into anisotropies in spin space and determines the easy axis of the magnetic ordered moment [Fig. 1(b)], and its connection with the
electronic nematic phase and superconductivity \cite{RMFernandes2014,Fernandes12}. Since a nematic quantum critical point is believed to occur near optimal superconductivity
in electron and hole-doped iron pncitides \cite{Kuo2016}, it is important to determine the temperature and electron/hole doping evolution
of SOC and its association with the nematic phase and superconductivity.  

One way to achieve this in iron pnictides is to study 
the energy, wave vector, temperature, and doping dependence of the 
spin excitation anisotropy using neutron polarization analysis. Compared with angle resolved photoemission experiments \cite{Borisenko}, polarized neutron scattering experiments  
typically have much better energy and momentum resolution \cite{dai}. 
In previous work on electron-doped BaFe$_{2-x}TM_{x}$As$_{2}$ \cite{Lipscombe,PSteffens,HQLuo2013,waer} and hole-doped Ba$_{1-x}$K$_{x}$Fe$_{2}$As$_{2}$ iron pnictides \cite{CZhang2013,NQureshi2014,YSong16}, 
there are clear evidence for spin excitation anisotropy in the paramagnetic tetragonal phase 
with $M_a\approx M_c>M_b$, where $M_a$, $M_b$, and $M_c$ are spin excitations polarized along
the $a$, $b$, and $c$-axis directions of the AF orthorhombic lattice, respectively, 
at temperatures well above $T_{N}$ and $T_s$ \cite{HQLuo2013,YSong16}. 
Although low-energy spin waves in the parent compound BaFe$_2$As$_2$ are also anisotropic in the orthorhombic AF ordered state
with $M_c>M_b>M_a$ \cite{NQureshi12,CWangPRX}, temperature dependence of the inelastic magnetic scattering at the 
AF ordering wave vector ${\bf Q}_{AF}={\bf Q}_1=(1,0,1)$ [Figs. 1(b) and 1(c)] and an energy transfer of $E=10$ meV changes from isotropic to anisotropic on cooling below $T_N$ \cite{NQureshi12}.
However, the energy scale of isotropic paramagnetic scattering at $E=10$ meV in BaFe$_{2}$As$_{2}$ is considerably larger than that of the anisotropic 
paramagnetic spin excitation in doped superconductors 
($E<6$ meV)  \cite{HQLuo2013,waer,CZhang2013,NQureshi2014,YSong16}. 
Since the SOC-induced spin space anisotropy is present in the paramagnetic tetragonal phase of doped iron pnictide superconductors and is also 
expected to be present in undoped BaFe$_2$As$_2$, it is possible that paramagnetic spin excitations in BaFe$_{2}$As$_{2}$ are also anisotropic, 
but with an energy scale smaller than $E=10$ meV.

To test if this is indeed the case, we carried out polarized neutron scattering experiments on BaFe$_{2}$As$_{2}$ with $T_N\approx T_s\approx 136$ K to study the 
temperature dependence of the spin excitation anisotropy [Fig. 1(d)].  In the AF ordered state at $T=135$ K, we find $M_c>M_b>M_a$ at ${\bf Q}_{AF}=(1,0,1)$ [Figs. 2(a), 2(b), and 3(a)], 
confirming the earlier results at 10 K \cite{NQureshi12,CWangPRX}. On warming to $T=138$ K ($>T_N,T_s$) in the paramagnetic tetragonal state, spin excitations at ${\bf Q}_{AF}=(1,0,1)$ are still
anisotropic below $E=4$ meV but with $M_a>M_b\approx M_c$ [Figs. 2(c), 2(d), and 3(b)]. For comparison, spin excitations at the AF zone boundary (ZB) 
${\bf Q}_{ZB}=(1,0,0)$ are isotropic for energies above $E=2$ meV [Fig. 3(d)].  Upon further warming to $T=160$ K, paramagnetic scattering becomes isotropic 
at all energies probed ($8\geq E\geq 2$ meV) [Fig. 3(c)].  While temperature dependence of the spin excitations at $E=8$ meV and ${\bf Q}_{AF}=(1,0,1)$ transforms from isotropic to anisotropic 
below $T_N$ with no evidence of critical scattering consistent with earlier measurements at $E=10$ meV \cite{NQureshi12}, paramagnetic scattering at $E=2$ meV  starts to develop spin space anisotropy below
about 160 K with enhanced $M_a$ ($>M_b\approx M_c$) on approaching $T_N$ due to condensation 
of the longitudinal component of the magnetic critical
scattering into $a$-axis aligned AF Bragg peak below $T_N$ [Fig. 4(a)-4(f)] \cite{Wilson10}.  On the other hand, paramagnetic scattering at $E=2$ meV  and ${\bf Q}_{ZB}=(1,0,0)$ is isotropic 
at all temperatures above $T_N$ [Fig. 4(g)-4(h)]. By comparing these results with spin excitation anisotropy  
in the paramagnetic tetragonal phase of electron/hole doped iron pnictide superconductors \cite{HQLuo2013,waer,CZhang2013,NQureshi2014,YSong16}, we conclude that electron/hole doping in 
BaFe$_{2}$As$_{2}$ necessary to induce superconductivity also enhances the 
$c$-axis polarized spin excitations associated with superconductivity. These results are 
also in line with the tetragonal $C4$ magnetic phase with spins aligned along the $c$-axis in near optimally hole doped  
superconducting Ba$_{1-x}$K$_{x}$Fe$_{2}$As$_{2}$ \cite{avci12,avci14,waer15,allred16}.

\begin{figure}[t]
\includegraphics[scale=.27]{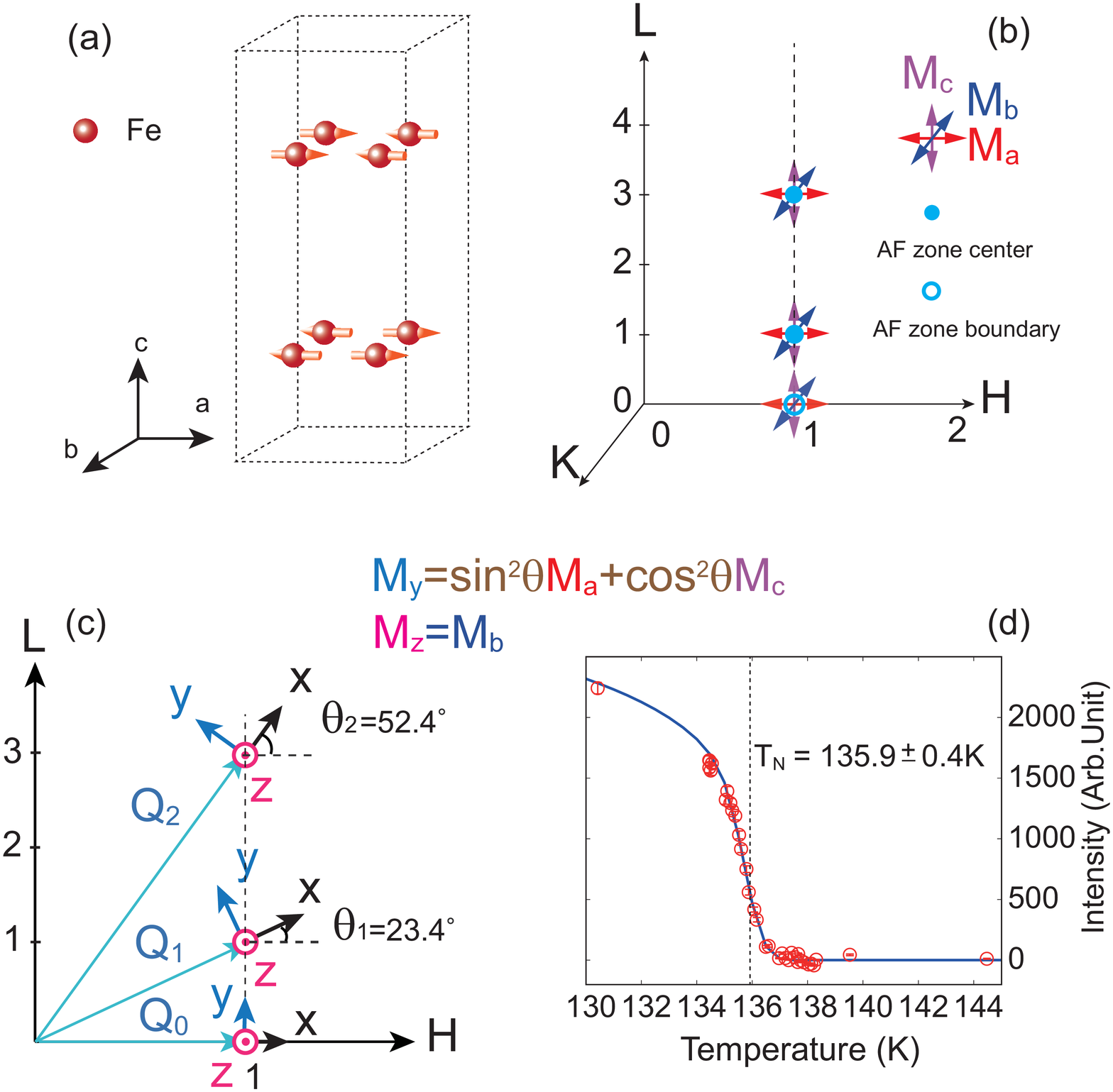}
\caption{(Color online) (a) The orthorhombic unit cell of BaFe$_2$As$_2$ (enclosed by dashed line) with only the magnetic Fe ions shown as red spheres. 
The arrows indicate the ordered moment direction along the longer $a$-axis.  Along the $c$-axis the nearest neighboring spins are antiparallel. 
(b) The positions of reciprocal space probed in the present experiment. Magnetic fluctuations polarized along the $a$, $b$, and $c$ directions are marked as 
$M_a$, $M_b$, and $M_c$, respectively.
 (c) Schematic of the $[H,0,L]$ scattering plane, where wave vectors ${\bf Q}_0$, ${\bf Q}_1$, and ${\bf Q}_2$ are probed. 
The neutron polarization directions are along the $x$, $y$, and $z$. The angle between the $x$ direction and $H$-axis is denoted as $\theta$. 
(d) The temperature dependence of magnetic order parameter measured at ${\bf Q}_1=(1,0,1)$. 
The solid line is a Gaussian convolved power law fit with $T_N=135.9\pm 0.4$ K.
}
\end{figure}

\begin{figure}[t]
\includegraphics[scale=.65]{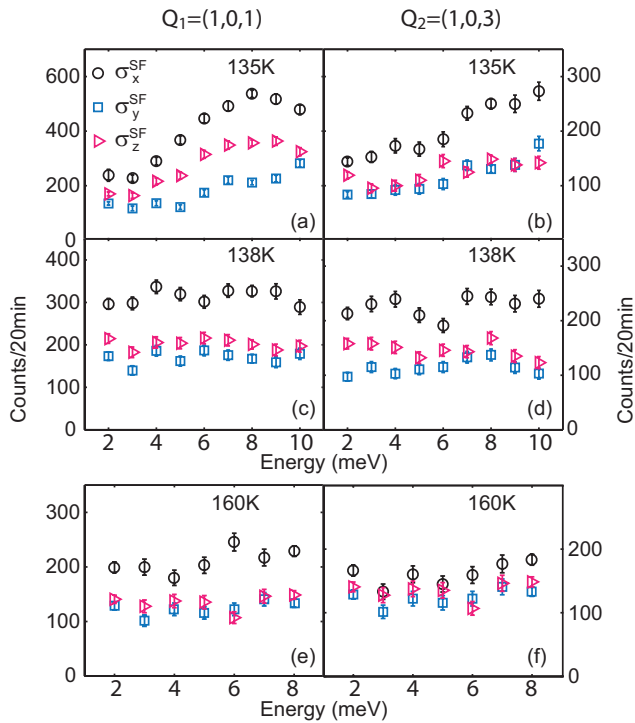}
\caption{(Color online) Constant-{\bf Q} scans at the two AF wave vectors ${\bf Q}_1=(1,0,1)$ and ${\bf Q}_2=(1,0,3)$ at $T=135$, 138, and 160 K. 
All three spin-flip (SF) channels $\sigma^{SF}_x$, $\sigma^{SF}_y$, $\sigma^{SF}_z$ are measured at these wave vectors as defined in Fig.1 (c).}
 \end{figure}

\begin{figure}[t]
\includegraphics[scale=.5]{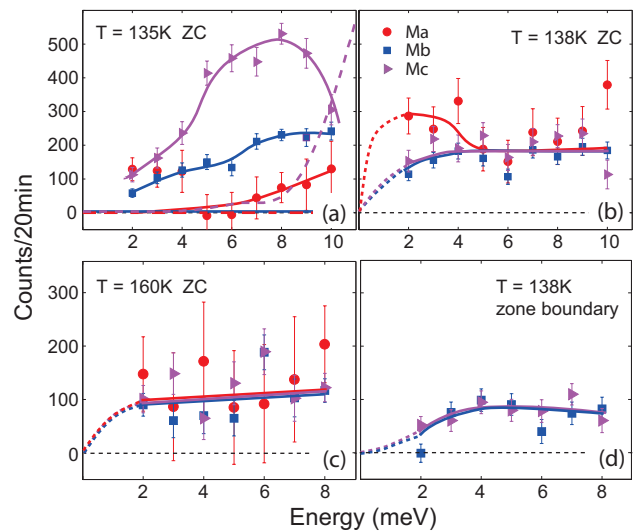}
\caption{(Color online) (a-c) The magnetic components $M_a$, $M_b$, and $M_c$ at ${\bf Q}_{AF}$ and 135 K, 138 K, and 160 K obtained from data in Fig. 2. 
The dashed lines in (a) are results for spin waves of BaFe$_2$As$_2$ at 2 K \cite{CWangPRX}. (d) The magnetic components at the zone boundary (ZB) at 138 K. 
Only $M_b$ and $M_c$ can be determined from measurements at ${\bf Q}=(1,0,0)$.}
\end{figure}

\begin{figure}[t]
\includegraphics[scale=.48]{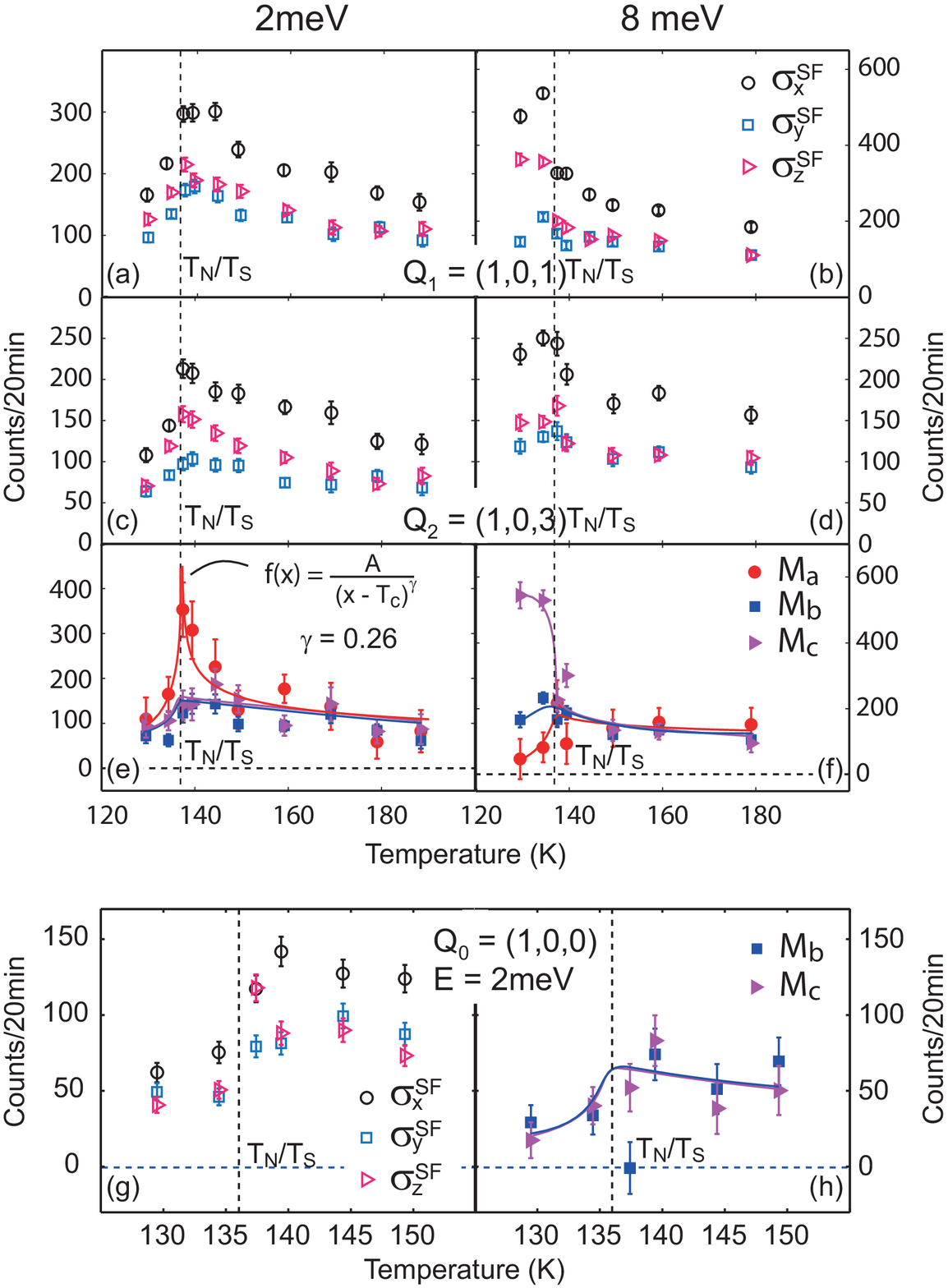}
\caption{(Color online) Temperature dependence of $\sigma^{SF}_x$,$\sigma^{SF}_y$, and $\sigma^{SF}_z$ at $E=2$ meV with (a) ${\bf Q}_1=(1,0,1)$ and (c) ${\bf Q}_2=(1,0,3)$. Similar data at $E=8$ meV with 
(b) ${\bf Q}_1=(1,0,1)$ and (d) ${\bf Q}_2=(1,0,3)$. (e) Temperature dependence of the components $M_a$, $M_b$, and $M_c$ at $E=2$ meV. 
The solid curve is the fitted line with function $f(T)=\frac{A}{(T-T_c)^\gamma}$. (f) Temperature dependence of $M_a$, $M_b$, and $M_c$ at $E=8$ meV. 
(g) The three SF scattering channels measured at ${\bf Q}_0=(1,0,0)$ and $E=2$ meV. (h) Temperature dependence of 
$M_b$ and $M_c$ determined from (g).The solid lines are guides to the eye.
The vertical dashed lines mark $T_N/T_s$.
}
\end{figure}

Our polarized neutron scattering experiments were carried out using the IN22 triple-axis spectrometers at the Institut Laue-Langevin, Grenoble, France. Polarized neutrons were produced using a focusing Heusler monochromator and analyzed with a focusing Heusler analyzer with a final wave vector of 
$k_f=2.662$ {\AA}$^{-1}$. About 12-g single crystals of BaFe$_2$As$_2$ used in previous work \cite{Haoran} are used in the present experiment. 
Figure 1(a) shows the collinear AF structure of BaFe$_2$As$_2$ with 
ordered moments along the $a$-axis \cite{qhuang,kim2011,Wilson10}.
The orthorhombic lattice parameters of the AF unit cell are $a\approx b\approx 5.549$ {\AA}, and $c=12.622$ {\AA}. The wave vector transfer ${\bf Q}$ in three-dimensional reciprocal space in {\AA}$^{-1}$ is defined as
${\bf Q}=H{\bf a^{*}}+K{\bf b^{*}}+L{\bf c^{*}}$,
with ${\bf a^{*}}=\frac{2\pi}{a}{\bf \hat{a}}$, ${\bf b^{*}}=\frac{2\pi}{b}{\bf \hat{b}}$
and ${\bf c^{*}}=\frac{2\pi}{c}{\bf \hat{c}}$, where $H$, $K$ and
$L$ are Miller indices. The samples were co-aligned in the $[H,0,L]$ scattering plane [Figs. 1(b) and 1(c)]. In this notation, the AF Bragg peaks occur at $[1,0,L]$ with $L=1,3,\ldots$, while the AF zone boundaries along the $c$-axis occur at $L=0,2,\ldots$. 
The magnetic responses at a particular ${\bf Q}$ along the $a$-, $b$-, and $c$-axis directions are marked as $M_a$, $M_b$, and $M_c$, respectively as shown in Fig. 1(b). In the paramagnetic tetragonal state, these correspond to magnetic excitations polarized along the in-plane longitudinal, in-plane transverse, and out-of-plane directions, respectively. 
The neutron polarization directions $x$, $y$, and $z$ are defined as along ${\bf Q}$, perpendicular to ${\bf Q}$ but in the scattering plane, 
and perpendicular to both ${\bf Q}$ and the scattering plane, respectively [Fig. 1(c)]. From the observed neutron spin-flip (SF) scattering cross sections $\sigma^{SF}_x$, $\sigma^{SF}_y$, and $\sigma^{SF}_z$, we can calculate the 
components $M_a$, $M_b$, and $M_c$ via $\sigma _x^{SF} = \frac{R}{R+1} (\sin ^2 \theta M_a + \cos ^2 \theta M_c) + \frac{R}{R+1} M_b +B$, 
$\sigma _y^{SF} = \frac{1}{R+1} (\sin ^2 \theta M_a + \cos ^2 \theta M_c) + \frac{R}{R+1} M_b +B$, and 
$\sigma _z^{SF} = \frac{R}{R+1} (\sin ^2 \theta M_a + \cos ^2 \theta M_c) + \frac{1}{R+1} M_b +B$, where $R$ is the flipping ratio 
($R=\sigma _{Bragg}^{NSF}/\sigma _{Bragg}^{SF}\approx 13$) and $B$ is the background scattering.  
By measuring $\sigma_{x,y,z}^{SF}$ at two equivalent AF zone center wave vectors ${\bf Q}_{AF}={\bf Q}_1 = (1,0,1)$ and ${\bf Q}_2 = (1,0,3)$, 
one can determine all three components of the magnetic response $M_a$, $M_b$, and $M_c$ \cite{HQLuo2013,waer,YSong16}. 
For the zone boundary position at ${\bf Q}_0=(1,0,0)$ with $\theta=0$, one can determine $M_b$ and $M_c$ using $\sigma_{x,y,z}^{SF}$ at this position.

To determine the magnetic ordering temperature of BaFe$_2$As$_2$, we show in Fig. 1(d) background subtracted 
 elastic SF cross section $\sigma _x^{SF}$ measured at ${\bf Q}_1 = (1,0,1)$.  The solid line is a fit of the magnetic order parameter  
with  Gaussian convolved power-law $M(T)^2=B^2\int(1-\frac{T}{T_N})^{2\beta}e^{-(T-T_N)^2/2\sigma^2}$ \cite{pajerowski}. Although this formula is used to account for sample inhomogeneities and a distribution of N$\rm \acute{e}$el temperatures 
in Co-doped Ba(Fe$_{1-x}$Co$_x$)$_2$As$_2$  \cite{pajerowski}, we use it for pure BaFe$_2$As$_2$, where disorder is not expected to be important, 
to compare with $\beta$ and $\sigma$ in lightly Co-doped samples.  We find $T_N=135.9\pm0.4$ K, $\sigma = 0.51 \pm 0.07$, and $\beta = 0.1 \pm 0.02$ for
BaFe$_2$As$_2$. While the value of $\sigma$ in BaFe$_2$As$_2$ is very similar to that of $x=0.021$ suggesting a small distribution of $T_N$ \cite{pajerowski}, 
the $\beta$ value is considerably smaller than the Co-doped samples but similar to previous value of $\beta=0.103$ for pure BaFe$_2$As$_2$ \cite{wilson09}.
Figure 2 shows energy scans at the AF wave vectors ${\bf Q}_1=(1,0,1)$ and ${\bf Q}_2=(1,0,3)$ at 
temperatures below and above $T_N$.  In an isotropic paramagnet with negligible background scattering and $R\rightarrow \infty$, we would expect
$\sigma _x^{SF} /2\approx \sigma _z^{SF} \approx \sigma _y^{SF}$.  
At $T=135$ K below $T_N$, 
magnetic scattering at ${\bf Q}_1=(1,0,1)$ shows strong anisotropy with $\sigma _z^{SF} > \sigma _y^{SF}$ [Fig. 2(a)].
Figure 2(b) plots similar scan at ${\bf Q}_2=(1,0,3)$ with $\sigma _z^{SF} \approx \sigma _y^{SF}$.
Since ${\bf Q}_1=(1,0,1)$ and ${\bf Q}_2=(1,0,3)$ correspond to angles of $\theta_1=23.4^\circ$ and $\theta_1=52.4^\circ$, respectively [Fig. 1(c)], 
we can use $\sigma_{x,y,z}^{SF}$ at these two wave vectors to completely determine $M_a$, $M_b$, and $M_c$ \cite{YS1,CL1}. 
Figure 3(a) shows our calculated $M_c$, $M_b$, and $M_a$ ($M_c>M_b>M_a$), and the outcome is similar to spin excitations 
of BaFe$_2$As$_2$ \cite{CWangPRX} and BaFe$_{1.91}$Co$_{0.09}$As$_{2}$ \cite{waer} in the low-temperature AF ordered phase.

In previous work, it was found that paramagnetic spin excitations of BaFe$_2$As$_2$ above $T_N$ and $T_s$ are isotropic at $E=10$ meV and ${\bf Q}_1=(1,0,1)$ \cite{NQureshi12}. To see if spin excitation anisotropy is 
present at $T=138$ K ($>T_N,T_s$) in the paramagnetic tetragonal state, we carried out constant-${\bf Q}$ measurements at ${\bf Q}_1$ [Fig. 2(c)] and ${\bf Q}_2$ [Fig. 2(d)].
Inspection of the figures finds clear difference in spin excitations ($\sigma _z^{SF} > \sigma _y^{SF}$) below about $E\approx 6$ meV at ${\bf Q}_2$.   
Figure 3(b) shows the energy dependence of $M_a$, $M_b$, and $M_c$ obtained by using the data in Figs. 2(c) and 2(d), revealing $M_a>M_b\approx M_c$ for energies below 6 meV.
Upon further warming the system to 160 K ($>T_N,T_s$), magnetic signal at ${\bf Q}_1$  [Figs. 2(e)] and ${\bf Q}_2$  [Figs. 2(f)]  becomes purely paramagnetic 
isotropic scattering in the energy region probed satisfying 
$(\sigma _x^{SF}-B)/2\approx (\sigma _y^{SF}-B)\approx (\sigma _z^{SF}-B)$. The energy dependence of $M_a$, $M_b$, and $M_c$ shown in Fig. 3(c) confirm 
the isotropic paramagnetic nature of the scattering. Figure 3(d) shows the energy dependence of $M_b$ and $M_c$ as obtained from constant-${\bf Q}$ scan 
at the zone boundary ${\bf Q}_0=(1,0,0)$, indicating isotropic paramagnetic scattering at energies probed.

Figures 3(a)-3(c) summarize temperature evolution of the estimated $M_a$, $M_b$, and $M_c$ at the AF zone center ${\bf Q}_{AF}$, obtained by using data in 
Fig. 2 after taking into account the magnetic form factor differences at ${\bf Q}_1$ and ${\bf Q}_2$ and other effects as shown in Ref. \cite{CL1}. 
In the AF ordered state at $T=135$ K ($\approx T_N-1$ K), the $M_c$ component dominates the spin excitation spectrum below 10 meV, followed by $M_b$ and $M_a$ [Fig. 3(a)]. 
For comparison, the $M_a$ component of the spin waves is completely gapped out below $\sim$10 meV at 2 K [dashed line in Fig. 3(a)]. 
When warming the system to $T=160$ K ($\approx T_N+24$ K), paramagnetic scattering is isotropic in spin space at all probed energies 
with $M_a=M_b=M_c$. At a temperature $T=138$ K ($\approx T_N+2$ K) slightly above $T_N$, paramagnetic spin excitations are anisotropic below $\sim$5 meV with $M_a>M_b\approx M_c$.

In previous unpolarized neutron scattering experiments on BaFe$_2$As$_2$ \cite{Wilson10}, two-dimensional (2D) magnetic critical scattering has been observed at temperatures far above 
$T_N$. Upon cooling, the longitudinal component of the critical scattering above $T_N$ ($M_a$) is expected to increase with decreasing temperature and condense into the
3D AF Bragg positions at the 2D-3D dimensional crossover temperature $T_{\rm 3D}$ near $T_N$ \cite{wilson09}.
The transverse components of spin excitations ($M_b$ and $M_c$) are the spin wave contributions not expected to diverge at $T_N$ \cite{Wilson10}.  To test if this is indeed the case, we measured temperature dependence of 
$\sigma_{x,y,z}^{SF}$ at $E=2$ meV and 8 meV at the AF zone center ${\bf Q}_1$ [Figs. 4(a) and 4(b)] and ${\bf Q}_2$ [Figs. 4(c) and 4(d)].
With decreasing temperature, $\sigma_{x,y,z}^{SF}$ increases in intensity with the differences between $\sigma_z^{SF}$ and $\sigma_y^{SF}$ most obvious near $T_N$ at ${\bf Q}_2$ [Fig. 4(c)].
Using data in Fig. 4(a) and 4(c), we estimate the temperature dependence of $M_a$, $M_b$, and $M_c$ in Fig. 4(e).  Consistent with the expectations from the magnetic critical scattering measurements \cite{Wilson10},
we see a diverging longitudinal spin excitations $M_a$ at $E=2$ meV while transverse spin excitations show no 
critical scattering around $T_N$.  On cooling below $T_N$,
all three polarizations of spin excitations are suppressed due to the formation of spin gaps \cite{NQureshi12}. 
Similar measurements at $E=8$ meV show isotropic paramagnetic scattering behavior ($M_a\approx M_b\approx M_c$) down to $T_N$ before splitting into $M_c>M_b>M_a$ seen in the
AF ordered state [Fig. 4(f)]. Figure 4(g) shows temperature dependence of the spin excitations  
$\sigma_{x,y,z}^{SF}$ at $E=2$ meV and zone boundary position ${\bf Q}_0$.  We see that magnetic scattering is isotropic at all measured temperatures with no evidence of spin anisotropy.

The diverging $M_a$ near $T_N$ in BaFe$_2$As$_2$ may arise from the longitudinally polarized spin excitation in the critical scattering regime of a Heisenberg antiferromagnet with Ising
spin anisotropy [Fig. 4] \cite{collin}.  This means that the effect of critical scattering in BaFe$_2$As$_2$ can force the fluctuating moment along the longitudinal ($a$-axis) direction
in the paramagnetic critical regime without the need for orthorhombic lattice distortion
and associated ferro-orbital (nematic) ordering. Although this scenario is interesting, we note that 
temperature dependence of spin excitation anisotropy in the paramagnetic state of AF ordered NaFeAs \cite{YSong13} and electron 
underdoped BaFe$_{1.904}$Ni$_{0.096}$As$_2$ \cite{HQLuo2013} behave differently.
 In previous polarized neutron scattering experiments on NaFeAs, 
which has a collinear AF order at $T_N=45$ K and an orthorhombic-to-tetragonal lattice distortion at $T_s\approx 58$ K \cite{Dinah,sli09}, 
$M_a\approx M_c$ is larger than $M_b$ in the paramagnetic orthorhombic phase below $T_s$ and the in-plane anisotropy $M_a-M_b$ enhances on approaching $T_N$ from $T_s$ \cite{YSong13}. 
When warming up to above $T_s$, the statistics of the data in NaFeAs is insufficient to establish possible spin anisotropy \cite{YSong13}.
Since one of the key differences between BaFe$_2$As$_2$ and NaFeAs is the coupled structural and magnetic phase transitions in BaFe$_2$As$_2$, our data 
suggest  
that the orthorhombic lattice distortion lifting the degeneracy of the Fe $d_{xz}$ and $d_{yz}$ orbitals also induces the $M_c$ and $M_b$ anisotropy.
This is consistent with 
the observation that $M_c$ has the lowest energy in spin waves of 
 the AF ordered BaFe$_2$As$_2$ \cite{NQureshi12,CWangPRX} and NaFeAs \cite{YSong13}, suggesting that it costs less energy for the $a$-axis ordered moment to rotate out of the plane than
to rotate within the plane.

For electron doped BaFe$_{1.904}$Ni$_{0.096}$As$_2$ 
superconductor with $T_c=19.8$ K and $T_N\approx T_s=33\pm 2$ K, spin excitation anisotropy at $E=3$ meV and zone center ${\bf Q}_{AF}$ with $M_a\approx M_c>M_b$
first appears below $\sim$70 K and shows no anomaly across $T_s/T_N$ before changing dramatically below $T_c$ \cite{HQLuo2013}. 
For hole-doped Ba$_{0.67}$K$_{0.33}$Fe$_2$As$_2$ superconductor with $T_c=38$ K and no structural/magnetic order, spin excitation anisotropy at $E=3$ meV and ${\bf Q}_{AF}$
with $M_a\approx M_c>M_b$ appears below $\sim$100 K, and also decreases abruptly $T_c$ \cite{YSong16}. 
The similarities of these results to those of NaFeAs in the nematic temperature regime ($T_s>T>T_N$) 
 suggest that the ferro-orbital order or fluctuations \cite{RMFernandes12,CCLee09,kruger,WCLv,CCChen,Valenzuela} in electron and hole-doped BaFe$_2$As$_2$ first appear in the 
paramagnetic tetragonal phase at temperatures well
above $T_s$ \cite{HQLuo2013,YSong16}.
Since SOC in iron pnictides is a single iron effect not expected to change dramatically as a function of electron and hole doping \cite{Cvetkovic,Fernandes14}, 
the weak/absence of $M_c$ and $M_b$ spin excitation anisotropy in the tetragonal phase of BaFe$_2$As$_2$ is difficult to understand.
One possibility is that the nearly coupled structural and magnetic phase transitions in BaFe$_2$As$_2$ \cite{kim2011} suppress the role of the SOC induced 
ferro-orbital fluctuations above $T_s$.  Although hole-doped Ba$_{1-x}$K$_{x}$Fe$_{2}$As$_{2}$ also has coupled structural and magnetic phase transitions in the underdoped regime \cite{avci12}, 
it changes to a double-{\bf Q} tetragonal magnetic
structure with ordered moments along the $c$-axis near optimal superconductivity \cite{avci14,waer15,allred16}.
When hole and electron doping in BaFe$_2$As$_2$ reduces the structural and magnetic ordering temperatures, the SOC induced 
ferro-orbital fluctuations start to appear at temperatures above $T_s$.  In this picture, the spin excitation anisotropy in the
superconducting iron pnictides originates from similar anisotropy already present in their parent compounds below $T_s$. The dramatic change in spin excitation
anisotropy across $T_c$ seen in electron and hole-doped BaFe$_2$As$_2$ suggests a direct coupling of the SOC to superconductivity. 
The systematic polarized neutron scattering measurements present here and in previous work on doped BaFe$_2$As$_2$ family 
of materials \cite{Lipscombe,PSteffens,HQLuo2013,waer,CZhang2013,NQureshi2014,YSong16} 
call for quantitative calculations on how SOC is associated with spin excitation anisotropy in iron pnictides.

The neutron scattering work at Rice is supported by the
U.S. NSF-DMR-1436006 and NSF-DMR-1362219 (P.D.). The materials synthesis efforts at
Rice are supported by the Robert A. Welch Foundation Grant
No. C-1839 (P.D.).

\end{document}